\documentclass{PoS}
\title{Sampling versus Blocking}

\ShortTitle{ Sampling versus Blocking  }

       \author{\speaker{Yannick Meurice}$^a$\footnote{Current email address: yannick-meurice@uiowa.edu},
       Yuzhi Liu $^{b}$, Judah Unmuth-Yockey$^a$, Li-Ping Yang$^c$, and 
Haiyuan Zou $^d$\\
\llap{$^a$} Department of Physics and Astronomy, University of Iowa, Iowa City, IA 52240, USA\\
\llap{$^b$} Department of Physics, University of Colorado, Boulder, CO 80309, USA\\
\llap{$^c$} Department of Physics,Chongqing University, Chongqing 400044, China\\
\llap{$^d$} Department of Physics and Astronomy, University of Pittsburgh, Pittsburgh, PA 15260, USA

}

\abstract{The idea of blocking in configuration space has played an important role in the development of the RG ideas. However, despite being half a century old and having had a huge intellectual impact, generic numerical methods to perform blocking for lattice models have progressed more slowly than sampling methods. Blocking may be essential to deal with near conformal situations. Typically, blocking methods have smaller statistical errors but larger systematic errors than sampling methods. This situation is evolving with recent developments based on the Tensor RG (TRG) method. We report recent results for spin and gauge lattice models obtained with this new method regarding searches for fixed points, calculations of critical exponents and resolutions of sign problems. An interesting model for comparison is the 2-dimensional O(2) model with a chemical potential which has a sign problem with conventional Monte Carlo but allows sampling with the worm algorithm and blocking with various TRG formulations. We compare the efficiency and accuracy of these two methods.} 

\FullConference{The 32nd International Symposium on Lattice Field Theory\\
                23-28 June, 2014\\
                Columbia University New York, NY}

\begin{document}

\section{Introduction}
The block spin idea has played a central role in the development of the RG method. However, successful applications of the method were possible, for instance the discovery of asymptotic freedom, without requiring successful numerical implementations  
of the original blocking idea. Blocking in configuration space is hard!  Anyone who believes that blocking is a straightforward procedure should attempt to write a simple algorithm for the two-dimensional Ising model on a square lattice.
For this model, blocking means replacing four spins in a 2x2  square block by a single variable and write an effective energy function (or at least some effective measure) for the new block variables. 
The procedure becomes more intricate at each step and finding the effective energy function is nontrivial. This is explained in more detail and illustrated in \cite{Exactblocking13prd}.

It is possible to invent approximations where no new interactions are generated by the blocking process. Well-known examples are the Migdal-Kadanoff approximation \cite{Migdal:1975zf}, the approximate recursion formula  or other hierarchical approximations \cite{hmreview}. However, in these examples, the lack of reference to an exact procedure makes the systematic improvement of these approximations difficult. 
In contrast, the TRG formulation allows us to write exact blocking formulas with numerically controllable truncations.  
The basic reason is that the TRG blocking separates neatly the degrees of freedom inside the block (which are integrated over), from those kept to communicate with the neighboring blocks.
The TRG approach of classical lattice models was introduced in Refs. \cite{JPSJ.64.3598} motivated by 
 tensor states developed in RG studies of quantum models.  Accurate truncation methods were used for the Ising model in Ref. \cite{PhysRevB.86.045139}. In Ref. \cite{Exactblocking13prd}, we showed that 
TRG methods can be applied to models studied by lattice gauge theorists, namely spin models ($O(N)$ and principal chiral models) 
and pure gauge models (Ising, $U(1)$ and $SU(2)$). The case of actions quadratic in Grassmann variables is briefly discussed in \cite{YMPRB13}. The cases of the $O(3)$ model \cite{judah} and Schwinger model \cite{schwinger} were discussed at this conference. 
\section{TRG blocking}
\label{sec:simple}
The first step in the TRG formulation of spin models is to use character expansions and assign degrees of freedom to the links. For the 2D Ising model, the partition function can be written exactly as
 \begin{equation}            Z = (\cosh (\beta))^{2V}\mbox{Tr} \prod_{i}T^{(i)}_{xx'yy'}\  ,
 \end{equation}
 where $\mbox{Tr}$ means contractions (sums over 0 and 1) over the links. 
 The horizontal indices $x,\ x'$ and vertical indices $y,\ y'$ take the values 0 and 1 and 
\begin{equation}
            T^{(i)}_{xx'yy'} =f_x f_{x'}f_y f_{y'} \delta\left(\rm{mod}[x+x'+y+y',2]\right) \ ,
            \label{eq:factor}
        \end{equation} where $f_0=1$ and $f_1 =\sqrt{ \tanh(\beta)}$. The delta symbol is 1 if $x+x'+y+y'$ is zero modulo 2 and zero otherwise. 
This representation reproduces the closed paths of the high temperature  expansion.  
 Blocking defines a new rank-4 tensor $T'_{XX'YY'}$, where each index now takes four values. Its explicit form is:
        \begin{eqnarray}
            \label{eq:square}
            &\ &T'_{X({x_1},{x_2})X'(x_1',x_2')Y(y_1,y_2)Y'(y_1',y_2')} = \\ \nonumber
            &\ &\sum_{x_U,x_D,y_R,y_L}T_{x_1 x_U y_1y_L}T_{x_Ux_1'y_2y_R}T_{x_Dx_2'y_R y_2'}T_{x_2x_Dy_Ly_1'}\  ,
        \end{eqnarray}
        where $X(x_2,x_2)$ is a notation for the product states e. g. ,
        $X(0,0)=1,\  X(1,1)=2, \  X(1,0)=3,\  X(0,1)=4$. 
 The partition function can now be written exactly as 
        \begin{equation}
            Z=(\cosh (\beta))^{2V}\mbox{Tr}\prod_{2i}T'^{(2i)}_{XX'YY'} \ , 
            \label{eq:ZP}
        \end{equation}
where $2i$ denotes the sites of the coarser lattice with twice the lattice spacing of the original lattice. 
The new expression has exactly the same form as the previous one except for the fact that the indices run over more values. 

In practice, this exact procedure will require truncations and we need to restrict the indices $x, y, \dots$ to a finite number $N_{states}$. 
We use the smallest blocking: $M^{(n)}_{XX'yy'}=\sum_{y''}T^{(n-1)}_{x_1x'_1yy''}T^{(n-1)}_{x_2x'_2y''y'}$
where $X=x_1\otimes x_2$ and $X'=x'_1\otimes x'_2$ take now $N_{states}^2$ values. 
We make a truncation $N_{states}^2\rightarrow N_{states}$ using
$T^{(n)}_{xx'yy'}=\sum_{IJ}U^{(n)}_{xI}M^{(n)}_{IJyy'}U^{(n)*}_{x'J}$. 
The unitary matrix $U$ diagonalizes a matrix which is either one of the following:
\begin{itemize}
\item
${\mathbb G}_{XX'}=\sum_{X''yy'}M_{XX''yy'}M^*_{X'X''yy'} $ (Ref.  \cite{PhysRevB.86.045139} )
\vskip10pt

\item 
${\mathbb T}_{XX'}=\sum_{y}M_{XX''yy}$ (Ref. \cite{YMPRB13}))
\end{itemize}
and we only keep the $N_{states}$ eigenvectors corresponding to the the largest eigenvalues of one of these matrices.
The overlap matrix $O_{ij}=\sum_XU_{iX}\tilde{U}^{*}_{Xj}$ of the two unitary matrices $U_{iX}$ and $\tilde{U}^{*}_{Xj}$ corresponding to the two different methods, appears to have remarkable properties. As shown in the $O(2)$ example below, the eigenvectors seem to be  approximately the same but the eigenvalues are sometimes in a different order. 
\vskip10pt


   \section{Recent results}
   For the Ising model on square and cubic lattices,  a truncation method  based on Higher Order SVD (HOSVD) gives accurate results for the 2D and 3D Ising models \cite{PhysRevB.86.045139}. The combination of TRG and HOSVD is abbreviated HOTRG hereafter. 
For the Ising model on a square lattice,  all truncation methods single out a surprisingly small subspace of dimension two \cite{YMPRB13}. 
In the two states limit, the transformation can be handled analytically yielding a value 0.964  for the critical exponent $\nu$ much closer to the exact value 1 than 1.338 obtained in Migdal-Kadanoff approximations.  Alternative blocking procedures that preserve the isotropy can improve the accuracy to $\nu=0.987$ (isotropic ${\mathbb G}$) and 0.993 (${\mathbb T}$ defined above) respectively.  
It was also observed that adding a few more states does not improve the accuracy \cite{efrati13}. 

The linear algebra seems insensitive to the fact that the  values of the initial tensor become complex.
This allows us to deal with complex $\beta$ or chemical potential (no apparent sign problem).
However, when one approaches a zero of the partition function, more states are necessary.
The TRG allows us to study the analyticity in complex $\beta$ and $\mu$ planes.
There are subtleties with parity at complex $\beta$ which may require CP or PT considerations. 
 For the Ising case, there is an excellent agreement with  the complex  Onsager-Kaufman exact solution and 
 the finite size scaling of  Fisher zeros of $O(2)$ agrees well with the Kosterlitz-Thouless scaling\cite{trgo2}.

 \section{Microscopic study of systematic errors}
\begin{figure}[h]
 \begin{center}
 \includegraphics[width=0.55\textwidth,angle=0]{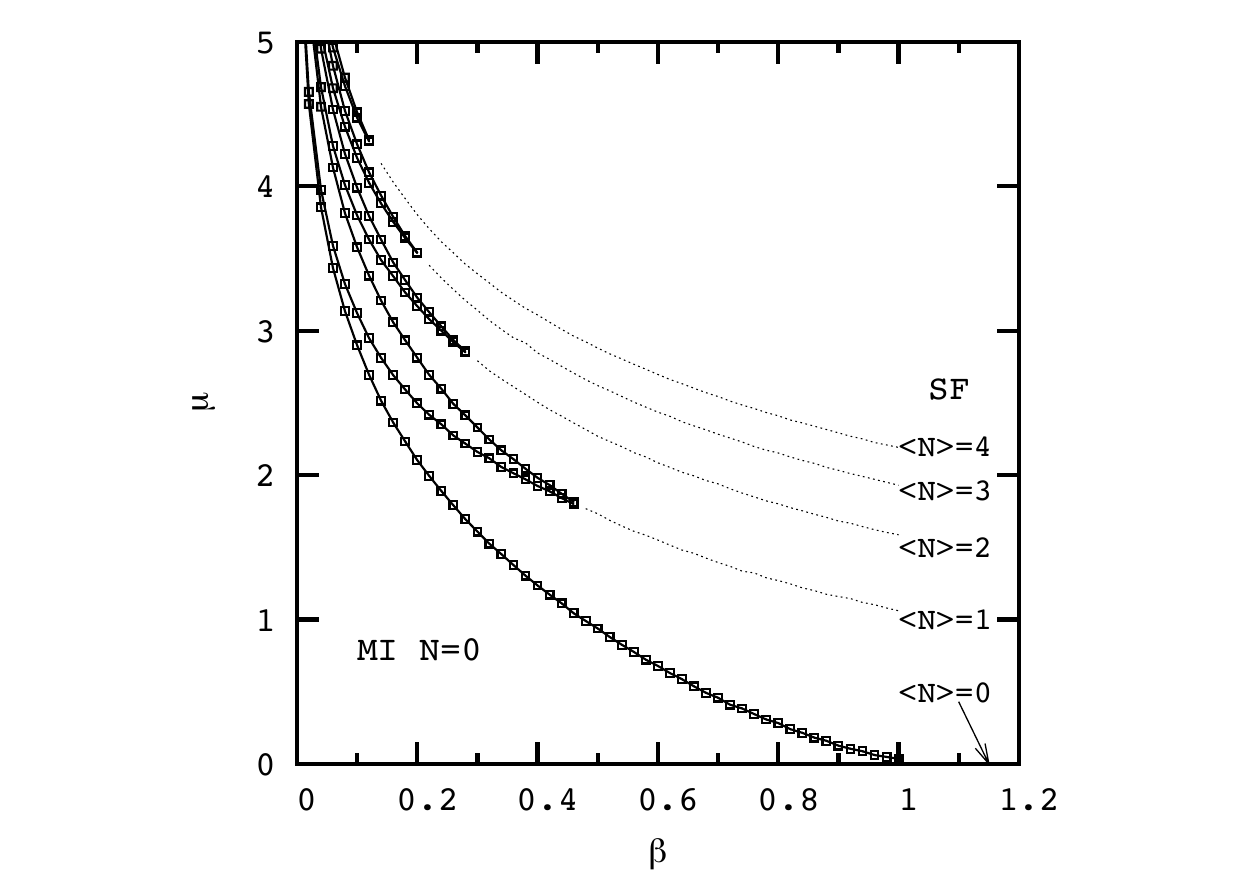}
  \includegraphics[width=0.35\textwidth,angle=0]{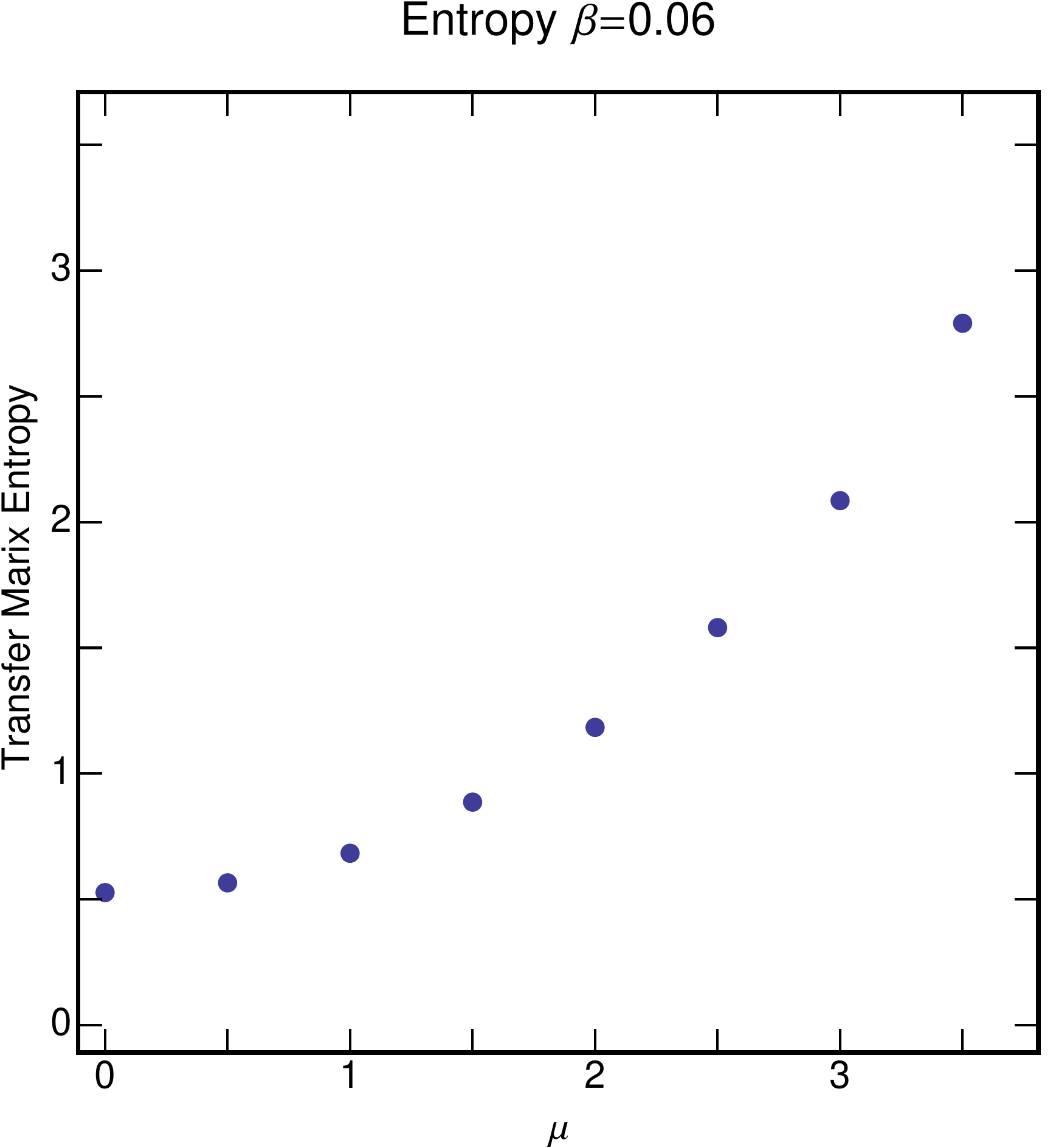}
\caption{\label{fig:phased}Phase diagram of the $O(2)$ model in the $\beta$-$\mu$ plane (left); Entropy of the transfer matrix defined below as a function of $\mu$ for $\beta$= 0.06 (right).}
\end{center}
\end{figure}

 Recently, we have used the TRG method to study  the $O(2)$ model with a chemical potential in one space and one Euclidean time direction  \cite{pra}. 
Our motivation was to suggest a testable way to do quantum simulations for this classical model using a many-body system that could be experimentally realized on 
an optical lattice. This model provides the simplest example of ``quantum rotors" used in this context. This question is not discussed here, but recent progress can be found in a separate contribution to this conference proceedings \cite{ab}. 
The $N_x\times N_\tau$ sites of the lattice are labelled $(x,t)$ and we assume periodic boundary conditions in space and time. The partition function reads:
\begin{eqnarray}
            Z &=& \int{\prod_{(x,t)}{\frac{d\theta_{(x,t)}}{2\pi}} {\rm e}^{-S}} \\
           S=&-&  \beta_\tau \sum\limits_{(x,t)} \cos(\theta_{(x,t+1)} - \theta_{(x,t)}+i\mu)\cr&-&\beta_s \sum\limits_{(x,t)} \cos(\theta_{(x+1,t)} - \theta_{(x,t)}).
 \end{eqnarray}
For $\mu \neq 0$ and real, the conventional MC method does not work (complex action). 
The partition function $Z$ can be calculated by taking the trace of $\mathbb{T}^{N_{\tau}}$ with $\mathbb{T}$ the transfer matrix. 
In \cite{pra}, we discuss the connection between the isotropic case ($\beta_s=\beta_\tau=\beta$) and the time continuum limit ($\beta_\tau>>\beta_s$), but here we focus on the isotropic case. 
We have used the TRG and the worm algorithm \cite{boris,Banerjee:2010kc} to calculate the particle number density
\begin{equation}
\left<N\right>\equiv \frac{1}{(N_s\times N_\tau)}\partial \ln Z /\partial \mu \ . 
\end{equation}
When we increase $\mu$ at fixed and not too large $\beta$, $\left<N\right>$ follows an alternated sequence of integer plateaus (the Mott insulating phase) and almost linear interpolation regions (the superfluid phase). This allowed us to construct the phase diagram in the $\beta$-$\mu$ plane as shown in the left part of Fig. \ref{fig:phased}. 

An important advantage of the TRG method is that it allows us to reach exponentially large volumes. However, it is important to check the results at small volume where sampling methods are feasible and accurate. 
The numerical values of $\left<N\right>$ 
on  a $16\times 16$ lattice, 
for values of $\beta_s=\beta_\tau$ and $\mu$ slightly below the 
tips of the  regions with $\left<N\right>=0,\ 1, \dots,\ 4$ of the phase diagram in Fig. \ref{fig:phased} are shown in 
Table \ref{tb:comp}.
\begin{table}[h]
\begin{center}
\begin{tabular}{|c|c|c|c|}
\hline
$\beta$ & $\mu$ & $\left<N\right>$ (worm) & $\left<N\right>$(HOTRG) \\ 
\hline
1.12 & 0.01 & 0.00726(1) & 0.00728(8) \\
0.46 & 1.8  & 0.98929(1)&0.9892(3)\\
0.28&2.85&1.98980(2)&1.989(2)\\
0.2&3.53&2.96646(3)&2.967(1)\\
0.12&4.3&3.96206(4)&3.965(1)\\
\hline
\end{tabular}
\end{center}
\caption{\label{tb:comp} Comparison of $\left<N\right>$ between worm algorithm and HOTRG}
\end{table}
It shows clearly that we have reached a very good agreement and the discrepancy appears at worse in the third significant digit. 
To the best of our knowledge the worm algorithm has only statistical errors and the accuracy can be improved by increasing the 
number of configurations. On the other hand,  the TRG has systematic errors associated with the truncation procedure that are important to understand ``microscopically" at each iteration. The results of this ongoing investigation will be published elsewhere \cite{shy}. 
We report here the preliminary results presented at the conference. The truncations were made with the transfer matrix rather than the HOSVD method mentioned above. No significant difference were seen between the two truncation methods. 
The overlap matrix $O_{ij}=\sum_XU_{iX}\tilde{U}^{*}_{Xj}$ discussed above, indicates  that the eigenvectors are approximately the same but the eigenvalues are sometimes in a slightly different order. An example of overlap matrix (for $\beta =0.06$, $\mu$=3.5; values smaller than $10^{-7}$ in absolute value have been replaced by 0) is provided here:
\[ \left( \begin{array}{cccccccc}

 1. & 0. & 0. & 0. & 0. & 0. & 0. & 0.  \\
 0. & 0. & 0.9983& 0. & 0. & 0. & 0.0576 & 0. \\
 0. & 0.9999 & 0. & 0. & 0. & 0. & 0. & 0. \\
 0. & 0. & 0. & 1. & 0. & 0. & 0. & 0. \\
 0. & 0. & 0. & 0. & 0.9997 & 0. & 0. & 0. \\
 0. & 0. & 0. & 0. & 0. & 1. & 0. & 0. \\
 0. & 0. & 0.0576 & 0. & 0. & 0. & 0.9983 & 0.  \\
 0. & 0. & 0. & 0. & 0. & 0. & 0. & 0.9996 
 \end{array} \right)\] 
 \vskip10pt

We started studying small systems in the middle of the superfluid phase between the $N$= 0 and 1 Mott insulating phases. We picked the values ($\beta =0.06$, $\mu$=3.5) and ($\beta=0.3$, $\mu$=1.8). 11 states for the initial tensor and then enough states  (among the 121 product states) in the first blocking is enough to stabilize $\left<N\right>$ with 5 digits. The good agreement and the number of states that needed to be kept out of the 121 is shown in Table \ref{tb:comp2}. The table makes clear that, as expected, when we increase the temporal size $N_\tau$, less states are needed. 
\begin{table}
\begin{center}
\begin{tabular}{|c|c|c|c|c|c|}
\hline
$N_x \times N_\tau$ &$\beta$ & $\mu$ & $\left<N\right>$ (worm) & $\left<N\right>$(HOTRG)&number of states \\
\hline
2 $\times$ 2 &0.06&3.5&0.69156&0.69155&31\\
2 $\times$ 4 &0.06&3.5&0.54080&0.54079&15\\
2 $\times$ 2 &0.3&1.8&0.61204&0.61204&34\\
2 $\times$ 4 &0.3&1.8&0.47929&0.47930&18\\
\hline
\end{tabular}
\end{center}
\caption{\label{tb:comp2} Comparison of $\left<N\right>$ between worm algorithm and HOTRG}
\end{table}

\begin{figure}[h]
  \includegraphics[width=0.32\textwidth,angle=0]{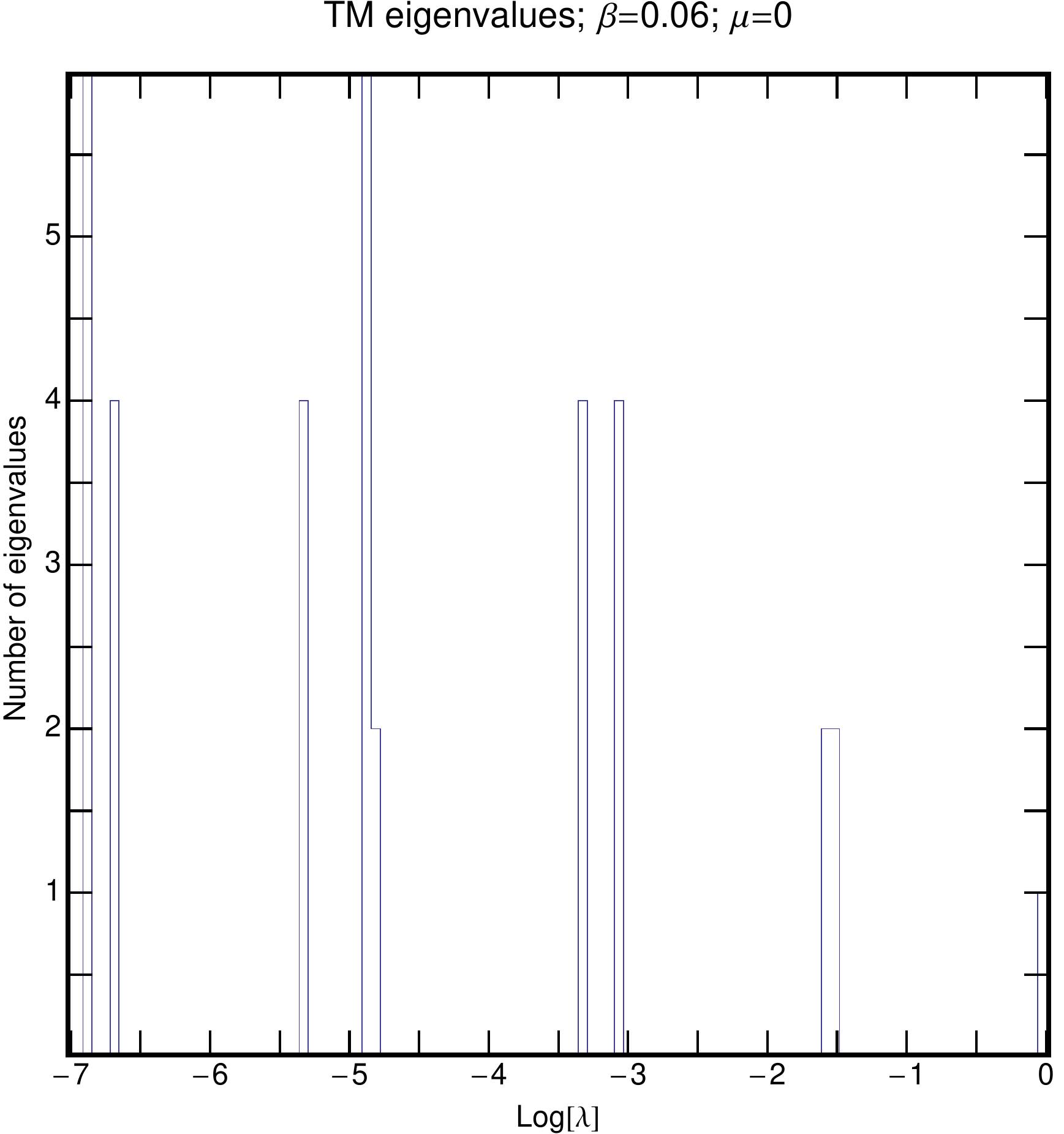}
  \includegraphics[width=0.32\textwidth,angle=0]{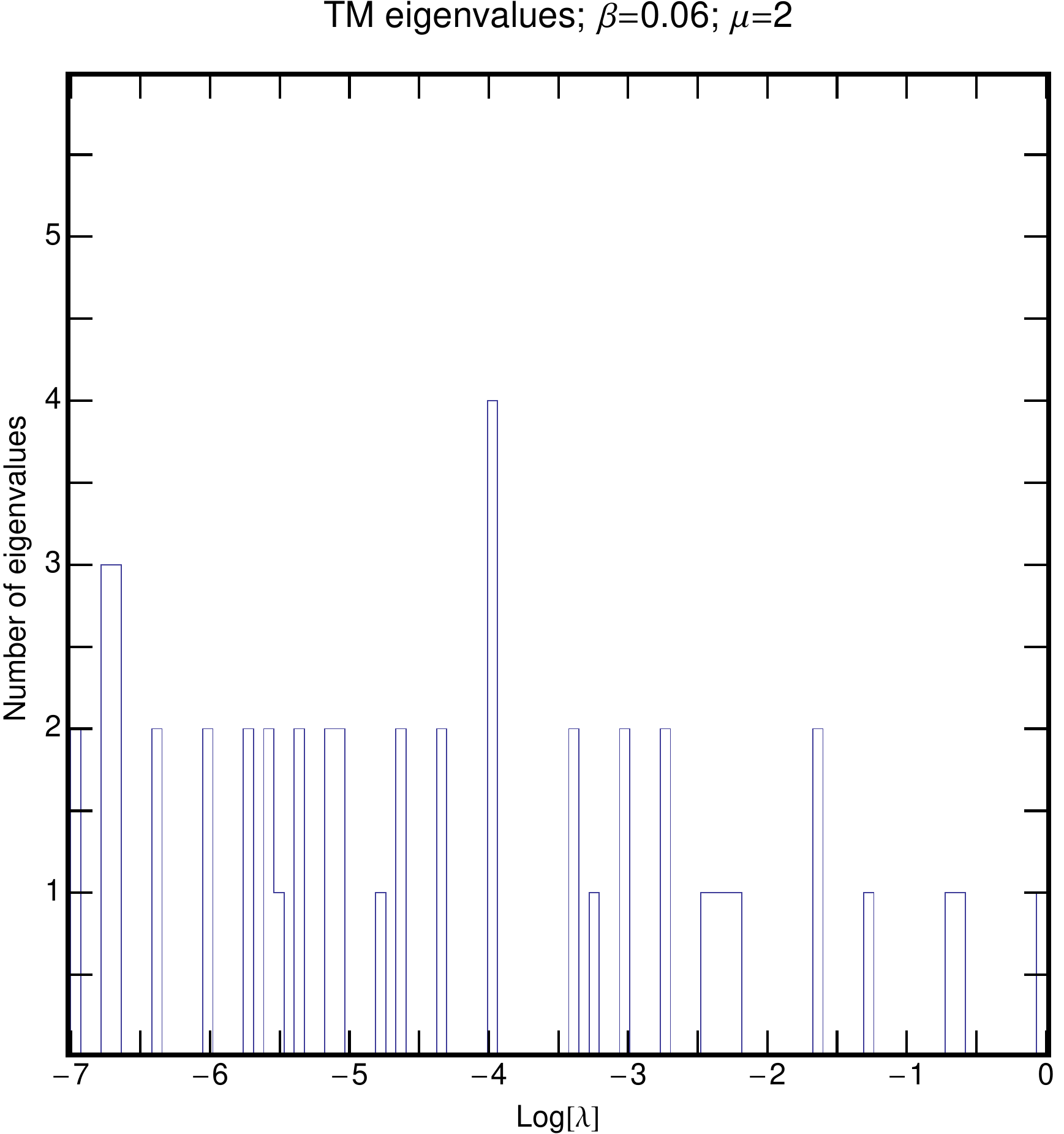}
  \includegraphics[width=0.32\textwidth,angle=0]{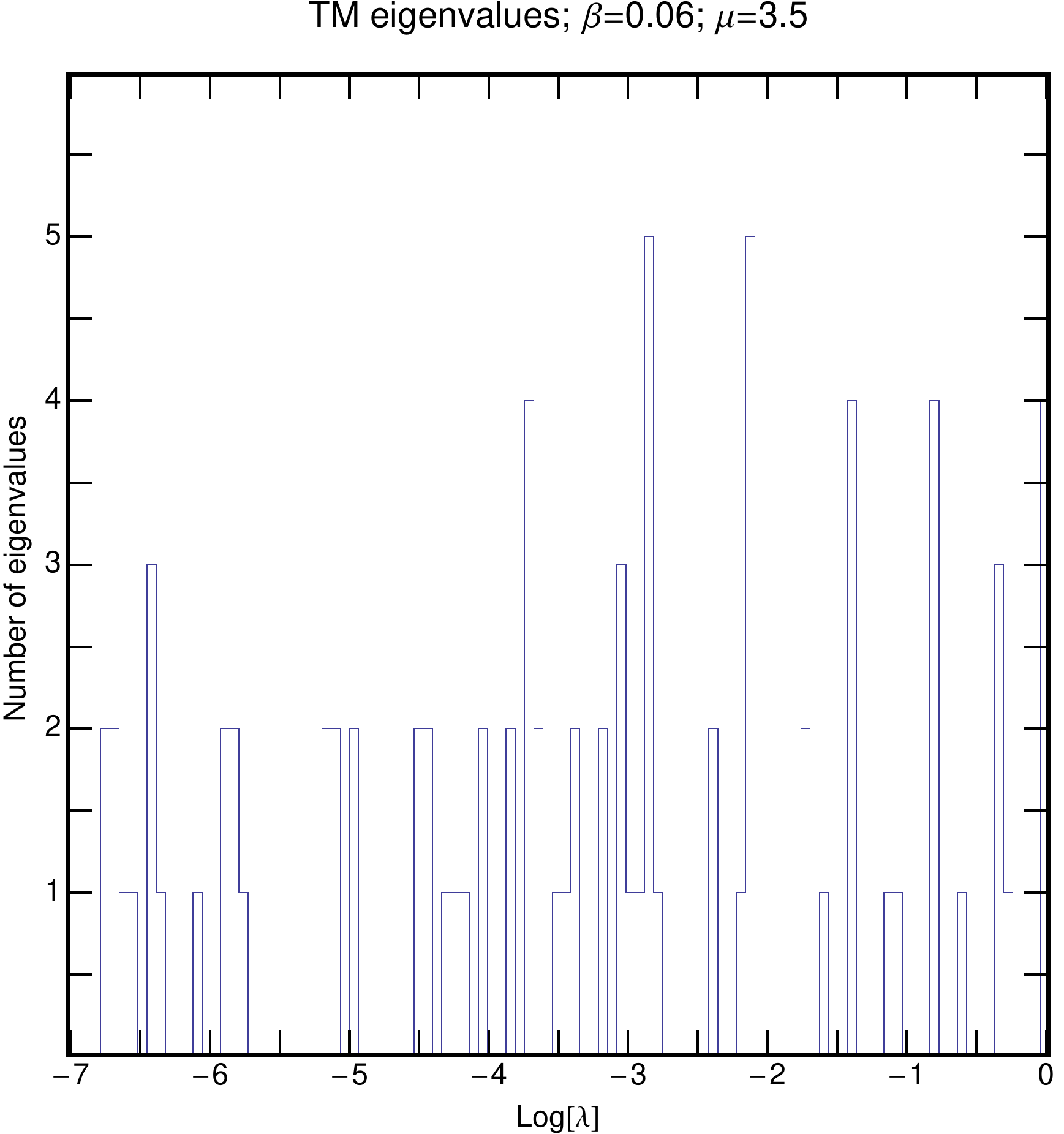}
\caption{ \label{fig:distr}Histograms of the logarithm of the eigenvalues of the transfer matrix for $\beta$ =0.06 and $\mu$ = 0, 2, and 3.5. The eigenvalues smaller than $10^{-7}$ are not displayed. }
\end{figure}

It is clear that the truncation becomes less accurate when the eigenvalues of the transfer matrix start populating the large value side. 
Increasing $\mu$ creates this tendency as shown in Fig.  \ref{fig:distr}. 
An indicator of this tendency is the entropy of the eigenvalue distribution. 
The eigenvalues of the transfer matrix are all positive, and  after normalization can be interpreted as probabilities, namely 
$\sum_i p_i =1$ . We can then define an invariant entropy $S=-\sum_i p_i \ln(p_i)$. The increase of this entropy with $\mu$ for $\beta=0.06$ is shown in the right part of Fig. \ref{fig:phased}. In the superfluid region, the entropy seems to 
increase with the iterations. When working on square lattices, this may be counteracted by the fact that the time direction becomes larger. A better indicator may be the entropy for the eigenvalues to the power $N_\tau$. This is work in progress \cite{shy}.
\section{Conclusions}

In summary, the TRG method provides exact blocking with controllable approximations.
 It deals well with sign problems and is reliable at larger Im$\beta$ than reweighting MC.
 For the Ising case, we found excellent agreement with  the complex  Onsager-Kaufman exact solution.
 The finite size scaling of  Fisher zeros of $O(2)$ agrees with Kosterlitz-Thouless.
We have robust estimations of the eigenvalues of the transfer matrix. We are in the process of understanding the 
very small discrepancies with the worm algorithm at the 5 digit level. The methods provides a connection between the Lagrangian and Hamiltonian approach. Our recursive construction of the transfer matrix provides the eigenvectors explicitly and could be used to 
calculate  real time evolution.

\vfill
\eject
\noindent
{\bf Acknowledgements}

We thank Boris Svistunov and other participants of the KITPC workshop ``Precision Many-body Physics of Strongly correlated Quantum Matter" for valuable conversations. 
We thank Shailesh Chandrasekharan for providing the worm algorithm code \cite{Banerjee:2010kc} used to compare the TRG calculations. 
 This research was supported in part  by the Department of Energy
under Award Number DE-SC0010114. 
We used resources of the National Energy Research Scientific Computing Center, which is supported by the Office of Science of the U.S. Department of Energy No. DE-AC02-05CH11231. Part of the simulations were done at CU Boulder Janus clusters.


\end{document}